\documentclass[prb,twocolumn,floatfix,showpacs,superscriptaddress]{revtex4-1}

\usepackage{amsmath}
\usepackage{amssymb}
\usepackage{amsfonts}
\usepackage{graphicx}
\usepackage{dcolumn}
\usepackage{bm}
\usepackage{color}

\DeclareGraphicsExtensions{.pdf,png,.jpg,.eps}

\begin{document}

\title{Interplay between symmetry and spin-orbit coupling in graphene nanoribbons}
\author{Hern\'an Santos}
\affiliation{Departamento de F\'{\i}sica Fundamental, Universidad Nacional de Educaci\'on a Distancia (UNED), Apartado 60141, E-28080 Madrid, Spain
}

\author{M. C. Mu\~noz}
\affiliation{Instituto de Ciencia de Materiales de Madrid (ICMM),
Consejo Superior de Investigaciones Cient\'{\i}ficas (CSIC),
C/ Sor Juana In\'es de la Cruz 3,
28049 Madrid, Spain}

\author{M. P. L\'opez-Sancho}
\affiliation{Instituto de Ciencia de Materiales de Madrid (ICMM),
Consejo Superior de Investigaciones Cient\'{\i}ficas (CSIC),
C/ Sor Juana In\'es de la Cruz 3,
28049 Madrid, Spain}

\author{Leonor Chico}
\affiliation{Instituto de Ciencia de Materiales de Madrid (ICMM),
Consejo Superior de Investigaciones Cient\'{\i}ficas (CSIC),
C/ Sor Juana In\'es de la Cruz 3,
28049 Madrid, Spain}

\date{\today }

\begin{abstract}  

We study the electronic structure of chiral and achiral graphene nanoribbons with symmetric edges, 
including curvature and spin-orbit effects. Curved
 ribbons show spin-split bands,  
whereas flat ribbons present spin-degenerate bands. 
We show that this 
effect is due to the breaking 
 of spatial inversion symmetry in curved graphene nanoribbons, while flat ribbons with symmetric edges possess an inversion center, 
 regardless of their having chiral or achiral edges.
We find an 
enhanced edge-edge coupling and a substantial gap in narrow chiral nanoribbons, 
which is not present in zigzag ribbons of similar width. 
We attribute these size effects to the mixing of the sublattices imposed by the edge 
geometry, 
 yielding a behavior of chiral ribbons that is distinct 
from those with pure zigzag edges.

\end{abstract}


\maketitle

\section{Introduction}\label{Introduction}

The crucial 
interplay between structure 
and electronic properties of
graphene is among the most attractive features of its derived
nanomaterials. Both carbon nanotubes (CNTs) and graphene nanoribbons 
(GNRs) show promising 
characteristics for spintronic devices.
Recent progress in experimental techniques has allowed for the fabrication 
of graphene nanostripes by using electron-beam
lithography\cite{CGPNG09} or by unrolling CNTs.\cite{KHS09} These ribbons could be used
in electronic devices, such as 
field-effect transistors,\cite{YVG11}
opening new perspectives 
for nanoelectronics.

The presence of localized edge states in GNRs, theoretically predicted,
\cite{FWNK96,WOTFN10}
and experimentally proved, \cite{Niimi_2006} confers them distinct properties.
GNRs have attracted a great amount of theoretical work, but 
mostly focused on
high-symmetry zigzag and armchair achiral ribbons. Zigzag ribbons have edges states with 
different spin polarizations, while armchair nanoribbons do not have edge states. 
The edges of minimal\cite{AK08,JA10} chiral ribbons can be considered as a mixture of armchair and zigzag edges, 
thus having edge-localized states stemming from their zigzag part. 
Although the evolution of the nanoribbon band structure upon the change of chirality
has been recently addressed,\cite{WTYS09,Waka_2012,YCL11} these systems have been nonetheless much less studied and many aspects remain to be explored.

The seminal work of Kane and Mele \cite{KM05} triggered the interest
on new quantum phases of matter and on the spin-orbit coupling (SOC) effect which,
although known to be small in graphene, gives rise to important physics. In particular, the quantum spin Hall (QSH) phase has been widely addressed.\cite{HK10,BGM12}

Curvature is known
to enhance spin-orbit interaction;
its importance 
in SOC effects has been theoretically 
investigated for 
the honeycomb lattice, especially for CNTs,\cite{ando00,CLM04, HHGB06,CLM09,ISS09,Chico_2012}
and experimentally confirmed.\cite{KIRM08} 
Hybridization of ${\pi}$ and ${\sigma}$
orbitals, decoupled in flat graphene, is enhanced by curvature and thus
SOC effects are bolstered.\cite{CLM09} The interplay between curvature
and SOC in GNRs has been mostly focused in achiral ribbons,
with highly symmetric zigzag and armchair edges.\cite{ZS09,LSM11, GPF11} 
 For zigzag GNRs, dispersionless edge bands in the flat geometry were found to become dispersive 
because of SOC effects. 
Both ${\pi}$ and ${\sigma}$ edge states remain 
spin-filtered in the curved geometry, still localized at the boundaries of the ribbon, albeit with an in-plane spin component and a localization length
larger than for the flat case.\cite{LSM11}
Recent experiments\cite{SCB09,KHS09,TJYC11}
on chiral GNRs obtained by unzipping CNTs show a reminiscent curvature. 
Scanning tunneling spectroscopy  
 measurements revealed the presence of one-dimensional edge states, with an energy splitting dependent on the width of the ribbon.\cite{TJYC11} 
By comparison with calculations employing a $\pi$-band model with a Hubbard term
the width dependence of the edge state gap was interpreted as a consequence of 
spin-polarized edge states.\cite{TJYC11,YCL11}
Hence, the study of curvature effects in GNRs is relevant from the experimental and theoretical viewpoint.

 In this paper we address the study of this ampler  
class of ribbons with chiral edges, focusing on the differences of SOC effects in flat and curved nanoribbons.
We summarize our main results as follows: \\
\noindent
(i) We find that the bands of both, chiral and achiral, flat ribbons with symmetric edges are at least twofold spin degenerate due to spatial inversion symmetry. Curving the ribbons breaks this symmetry, thus yielding spin-split bands except for the time-reversal protected special symmetry $k$ points. \\
\noindent
(ii) We find a gap in all chiral ribbons, despite the fact that they have a zigzag edge component.  
Boundary conditions in chiral ribbons mix both sublattices at each edge. This enhances edge-edge coupling, which results in a substantial gap without invoking electron interactions.\\
\noindent
(iii) Curvature augments spin-orbit effects in GNRs, yielding a larger splitting in the spin-split bands. In fact, curvature may induce metallicity in ribbons which have a gap in the planar form. \\
\noindent 
(iv) The spatial distribution of edge states depends on curvature and chirality. While zigzag ribbons are known to have spin-filtered states at the edges, in narrow ($\approx 40$ \AA) chiral ribbons edge states can have nonzero density at both edges, due to the edge-edge coupling. This size effect is more evident in ribbons with chiral angle close to $30^{\rm o}$, i.e., that of the armchair edge, for which the sublattice mixing is stronger.

This paper is outlined as follows. Section \ref{sec:geom} describes the structure and geometry
of the ribbons studied. 
Section \ref{sec:sym} gives some symmetry considerations concerning the role of spatial inversion in flat and curved general ribbons. 
 Section \ref{sec:method} contains the description
of the model Hamiltonian and calculation method.  Section V presents 
the  results, including spin-orbit interaction, 
for ribbons of different widths in flat and curved geometries. Finally, 
 in Section VI we discuss our results and 
 final conclusions are drawn.

\section{Geometry}\label{sec:geom}

\begin{figure}[htbp]
\includegraphics*[width=80mm]{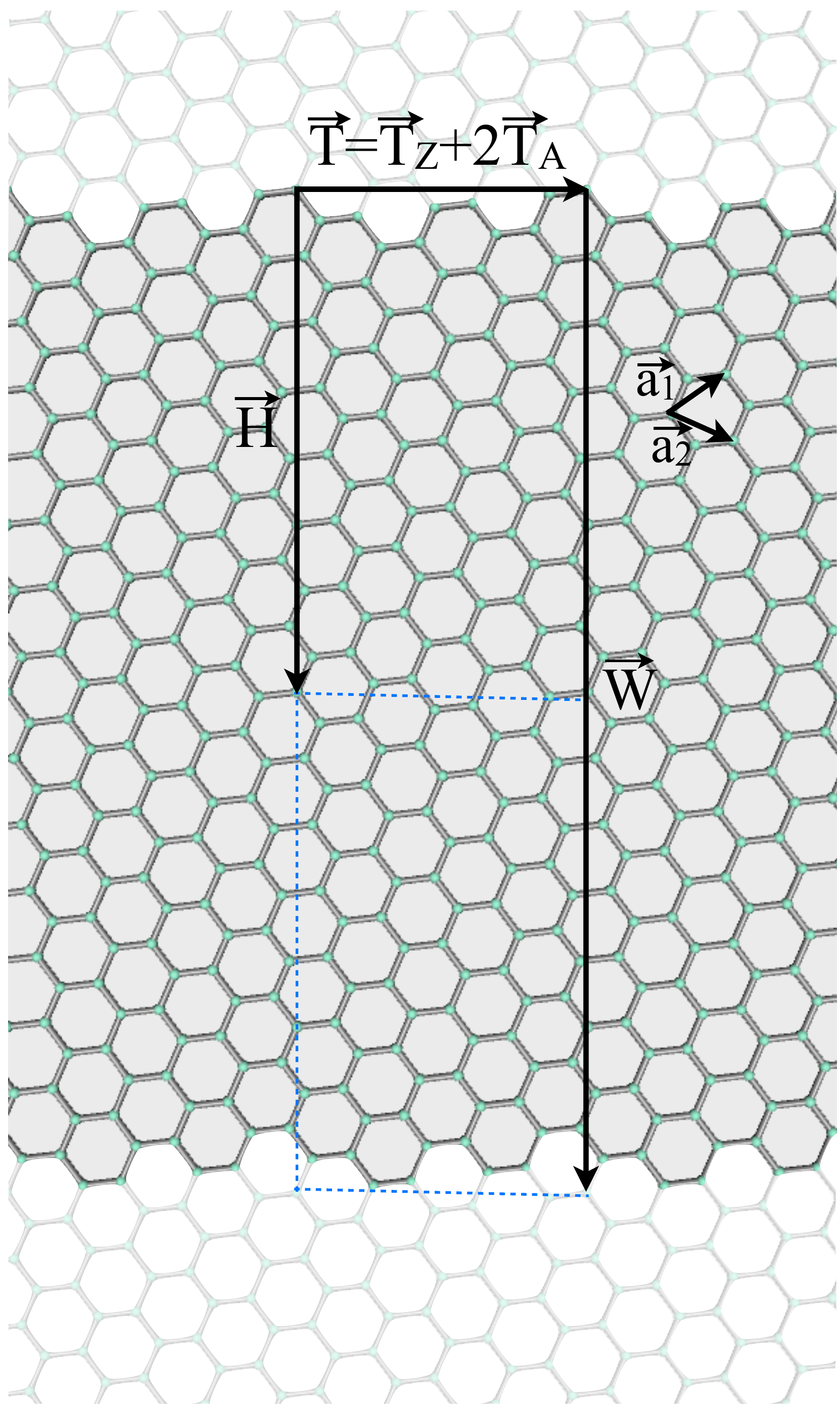}
\caption{(Color online) Geometry of the 2(3,2) GNR highlighted in dark gray on a graphene sheet, 
showing its translation vector ${\bm T}=  {\bm T}_Z + 2{\bm T}_A$ and its width 
vector ${\bm W}=  2{\bm H}$, where  ${\bm H}=  -7{\bm a}_1 + 1{\bm a}_2$.  
The unit cells spanned by ${\bm T}$ and ${\bm H}$ or ${\bm W}$ are indicated with dotted lines.}
 \label{fig_geom}
\end{figure}

We focus on chiral ribbons with symmetric minimal edges,\cite{AK08}  
obtained from unrolling chiral carbon nanotubes. 
The ribbon is thus characterized by the edge 
vector ${\bm T}=n{\bm a}_1+m{\bm a}_2$, where ${\bm a}_1$ and ${\bm a}_2$ 
are the primitive vectors of graphene, and the width  vector ${\bm W}$. 
The widths considered are therefore given by an integer multiple of ${\bm H}$, 
defined as the smallest graphene lattice vector perpendicular to ${\bm T}$, as 
depicted in Fig.\ref{fig_geom}. For a given ${\bm T}$, ${\bm H}$ is uniquely 
determined up to a global $\pm 1$ factor. As  ${\bm W} = M{\bm H}$, we will 
denote the ribbons by $M(n,m)$, where $M$ states the width of the ribbon, and $(n,m)$ 
indicates the minimal edge. All minimal edges can be decomposed in 
a zigzag and an armchair 
part,\cite{AK08,JA10} ${\bm T}=  n_Z{\bm T}_Z + n_A {\bm T}_A$, 
with  ${\bm T}_Z = {\bm a}_1$ and ${\bm T}_A={\bm a}_1+{\bm a}_2$.

The chirality of the ribbon is specified by the chiral angle ${\theta}$  between 
 the translation vector $(n,m)$ which defines the edges and the zigzag direction (1,0). 

We take into account  
different curvatures in the transversal direction for a given flat GNR, with no stretching allowed.  
 Curvature is denoted by 
the angle $\varphi$ spanned by the ribbon from its curvature center, ranging from zero for a flat ribbon to 
a value of 2$\pi$, 
which corresponds 
to a nanotube with cut bonds along its length. The degree of curvature is controlled by the angle
and by the diameter of the cylindrical configuration. 

\section{Symmetry considerations}\label{sec:sym}

\begin{figure}[htbp]
\includegraphics*[angle=0,width=80mm]{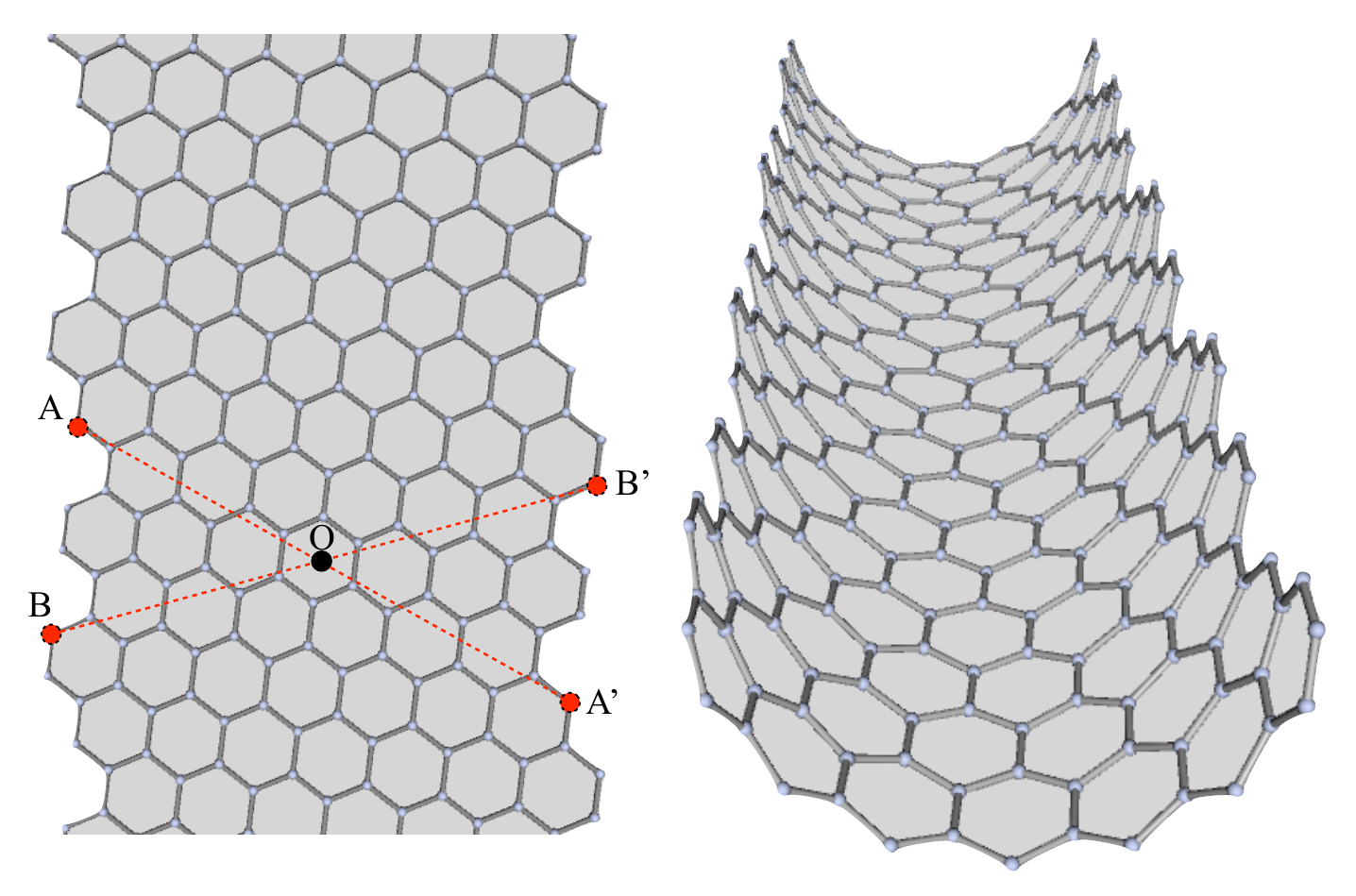}
\caption{(Color online) Schematic geometry of a flat (left) and curved (right) 1(3,2) GNR. The curvature angle is $\varphi=\pi$. Two pairs of equivalent atoms under spatial inversion symmetry are highlighted, and the inversion center is marked as $O$. }
 \label{fig_curv}
\end{figure}

Carbon nanotubes are classified as achiral and chiral according to their having a symmorphic or non-symmorphic symmetry group, respectively. This means that chiral tubes possess an spiral symmetry, so that there are two enantiomers for each chirality, while achiral tubes are equal to their mirror image; i.e.,  achiral tubes present space inversion symmetry while 
chiral tubes do not.\cite{SDD98,CBR07} 
Graphene nanoribbons, like their siblings carbon nanotubes,
are customarily classified as achiral and chiral according to their edge shapes.
In GNRs this classification is related to the
 crystallographic orientation of the boundaries: ribbons with zigzag and 
armchair edges (derived from armchair and zigzag CNTs, respectively) are called achiral, 
and those obtained from chiral tubes are called chiral GNRs. 
However, these so-called chiral ribbons with symmetric edges 
do have an inversion center. Upon bending the ribbon the
inversion symmetry is lost. This feature is crucial when considering
SOC effects. 
Fig. \ref{fig_curv} shows an example of a flat (left panel) and curved (right panel) unit cell
of the (3,2) ribbon, the latter with $\varphi=\pi$. A symmetry center is indicated
in the planar geometry. No such inversion center exists in the curved nanoribbon.

Notice that a flat nanoribbon with different edges lacks inversion symmetry. This situation is of experimental interest: most likely, actual ribbons will not have symmetric edges. Such asymmetry can be achieved either by adding or removing atoms to an originally edge-symmetric ribbon, or by altering the bond lengths in one of the edges. Dissimilar bond lengths may arise as a result of a different functionalization on the two edges of the nanoribbon.

\section{Theoretical model and computational methods}\label{sec:method}

We calculate  the band structure of graphene within the empirical
tight-binding (ETB) approximation. Although the $\pi$-orbital tight-binding model
is known to capture the low energy physics of graphene, since we are 
interested in SOC effects, we consider here
an orthogonal four-orbital $2s$, $2p_x$, $2p_y$, $2p_z$ basis set.
This allows for the inclusion of the intrinsic SO terms within the
conventional on-site approach.
The matrix Hamiltonian is built following the Slater-Koster formalism
up to nearest-neighbor hopping. We use the parametrization
obtained by Tom\'anek-Louie for graphite.\cite{TL88}
The expression of the one-electron ETB Hamiltonian is 
\begin{equation}
H_{0}= \sum_{i,\alpha,s} \epsilon_{\alpha}+\sum_{\langle i,j \rangle,\beta,s} t_{ij}^{\alpha,\beta} c_ {i, s}^{\alpha+}c_{j,s}^{\beta}+ H.c. ,
\end{equation}

\noindent
where $\epsilon_\alpha$ represents  the atomic energy of the
orbital $\alpha$, $\langle i,j \rangle$ stands for all the atomic sites of the unit
cell of the GNR, 
and $c_{i, s}^{\alpha+}$ and $c_{i, s}^{\alpha}$ are the
creation and annihilation operators, respectively, of one electron at site $i$,
orbital $\alpha$, and spin $s$.
We focus on neutral graphene; thus, no doping effects are addressed.

\noindent

SOC effects are included by adding an atomic-like term $H_{SO}$ to the 
$H_{0}$ Hamiltonian.
Assuming that the most important contribution of the crystal potential
to the spin-orbit coupling is close to the cores, the
$H_{SO}$ contribution takes the form 
$${ H_{SO}} = \sum_{i} \frac{\hbar}{4m^2c^2} \frac{1}{r_i} \frac{d V_i}{d r_i} {\bm L}\cdot{\bm  S} = \lambda {\bm L}\cdot{\bm S},$$

\noindent
where the spherical symmetry of the atomic potential $V_i$ has been assumed
and $r_i$ is the radial coordinate with origin at the $i$ atom.
${\bm L}$ stands for the orbital angular momentum of the electron, and
${\bm S}$ is the spin operator. The parameter $\lambda$ is a renormalized 
atomic SOC constant which depends on the orbital angular momentum.
Notice that  the $ H_{SO} $ terms only couple $p$
orbitals in the same atom. Considering the spin parts of the wave functions, 
the Hamiltonian matrix has $8N_a\times 8N_a$ elements, $N_a$ being the number
of the C atoms in the unit cell of the GNR and 8 corresponding to the
four orbitals per spin of the $sp^3$ basis set. 
The total Hamiltonian 
in the $2 \times 2$ block spinor structure is given by 

\begin{equation}
H = \left(
\begin{array}{cc}
H_0 + \lambda L_z & \lambda (L_x-iL_y)\\
\lambda (L_x+iL_y) & H_0 - \lambda L_z \\
\end{array}
\right).
\end{equation}

The total Hamiltonian $H$ incorporates both, spin conservation
and spin-flip terms. The spin-conserving diagonal terms 
act as an effective Zeeman field producing gaps at the
$K$ and $K'$ points of the graphene Brillouin zone (BZ), with opposite signs.\cite{CLM09}
By exact diagonalization of the matrix $H$ we obtain the
band structure of GNRs. 
As explained in the previous section, the curved geometry is obtained by
 isotropically 
bending the ribbon in the width direction, without changing the
distance along its length. Thus, no bond stretching is included along the ribbon axis. 
We do not consider reconstruction or relaxation of the edges or passivation
of the dangling bonds.

The value of the SOC constant for C-based materials is not well established and it is still under debate. 
Some theoretical estimates gave $\lambda=1\mu$eV for graphene,\cite{Min06,ZS07} much smaller than the atomic SO coupling, 8 meV.
 Taking into account the role of $d$ orbitals, this value raises to $\lambda=25\mu$eV.\cite{Konschuh10} 
 Accurate measurements of SOC are difficult to perform in graphene because external effects such as substrates, electric fields, or impurities may mask its value. 
 However, recent experiments in CNT quantum dots have reported  spin-orbit splittings substantially higher than those theoretically predicted: Kuemmeth {\it et al.} \cite{KIRM08} give a maximum splitting of 0.37 meV in a CNT of diameter 5 nm. One possible explanation for this energy splitting is that a higher value of $\lambda$ should be considered; as indicated by Izumida {\it et al.},\cite{ISS09} those measurements are compatible with $\lambda=14$ meV. More recently, Steele {\it et al.} \cite{Steele13} have presented evidence of large spin-orbit coupling in CNTs, up to 3.4 meV, an order of magnitude larger than previously measured and the largest theoretical estimates.  Furthermore, transport experiments report spin-relaxation times in graphene 1000 times lower than predicted.\cite{Lundeberg13} This is compatible with a larger value of the SOC coupling than those given by previous theoretical estimates.\cite{Min06,ZS07,Konschuh10} Although small, its effects in GNRs could have important consequences
when considering the spin degree of freedom, as has been experimentally shown in CNTs.\cite{KIRM08,Steele13} For the sake of clarity, we choose for the Figures a spin-orbit 
interaction parameter $\lambda=0.2$ eV. 

%

The spin-orbit contribution to the Hamiltonian, $ H_{SO} $ is linear on  $\lambda$. We have checked that, for small values of this parameter, such as the one employed here and those of physical relevance, the eigenvalues of the full Hamiltonian $H$ are basically a linear correction to those without the SOC term, $H_0$. Therefore, the spin-orbit splittings are proportional to $\lambda$ and the results presented in this work can be scaled accordingly. 


\section{Results}\label{sec:results}

We have calculated the electronic properties of many different chiral GNRs, varying their width and curvature. 
All calculations have been performed with the four-orbital parametrization explained in Sec. \ref{sec:method}. Therefore, they show some differences with respect to the widely used one-orbital approach.\cite{YCL11} 
We present here the 
band structures 
for three representative 
chiral ribbons [$M(7,1)$, $M(5,2)$, and $M(3,2)$] and, for comparison, some zigzag GNRs of different widths. The $M(7,1)$ ribbon 
has a chiral angle ${\theta=6.58^{\rm o}}$, close to the zigzag direction; the $M(5,2)$ has ${\theta=16.02^{\rm o}}$; 
and the $M(3,2)$ GNR has ${\theta=23.41^{\rm o}}$, closer to the armchair direction.
The $M(7,1)$ and $M(3,2)$ GNRs have the same unit cell with 76 atoms,
but with different orientation; i.e., 
the $\bm H$ and $\bm T$ vectors are interchanged. The $M(5,2)$ ribbon, in the
intermediate chirality range, has a unit cell with 52 atoms. The (5,2) edge has three armchair ($A$) and  two zigzag ($Z$) units, 
so for the infinite system there are two possible arrangements of the armchair and zigzag
units with the same edge vector. We choose the one with all zigzag units together, the $ZZZAA$. 
While the edge states in a semi-infinite graphene sheet or in very wide ribbons are the same irrespectively
 of the sequence, for narrow ribbons some differences in the band structure of distinct edge arrangements may arise.

\subsection{SOC and inversion symmetry}
\subsubsection{Curvature effects}

\begin{figure}[htbp]
\includegraphics*[clip,width=\columnwidth]{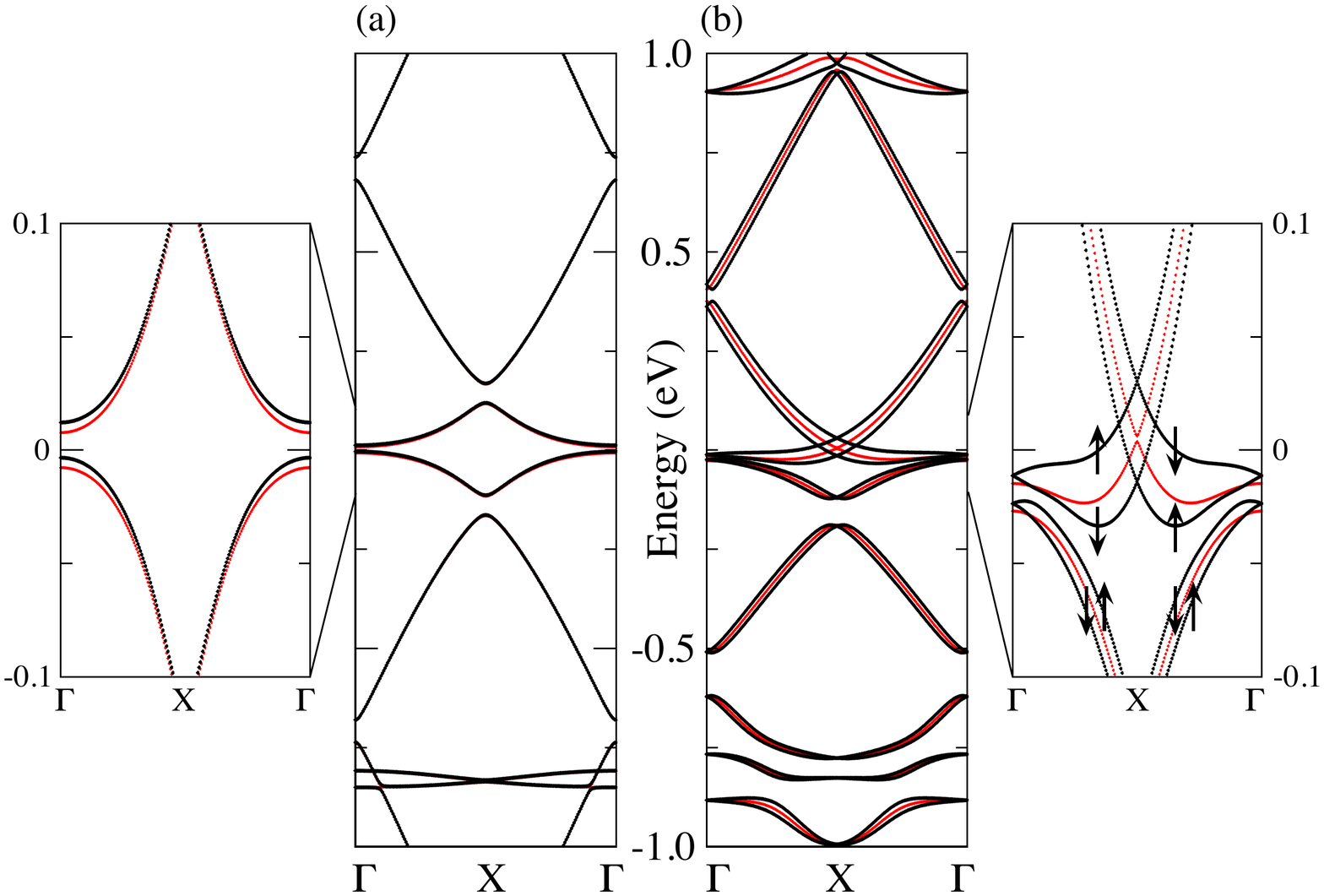}
\caption{(Color online) Electronic structure of
the 1(7,1) GNR
 calculated with (black dots) and without (red/gray dots) SOC, considering (a) the four-orbital
$sp^3$ basis set in a planar geometry; and (b) the same $sp^3$ basis but for a curved geometry with $\varphi=2\pi$. Zooms of the edge bands are included close to each panel. 
}
\label{3_71}
\end{figure}

In systems with time-reversal and spatial inversion symmetry, spin-orbit interaction does not lift spin degeneracy, according to Kramers' theorem. 
If spatial inversion symmetry is not present, the states are spin-split except in the $k$ points protected by time reversal invariance. 
In symmetric-edge GNRs, spatial inversion symmetry is broken 
by curving the ribbon, as indicated in Fig. \ref{fig_curv}. 
The importance of the broken inversion symmetry is shown in Fig. \ref{3_71}, where the electronic structures of the 1(7,1) ribbon calculated with and without SOC terms are depicted for (a) the planar configuration 
and (b) the curved one with $\varphi=2\pi$. This angle corresponds to 
a maximally curved geometry without overlapping the edges of the ribbon; it is equivalent to an open carbon nanotube with 
circumference equal to the width of the GNR. 
SOC effects are clear:
all degeneracies, including spin, are lifted in the curved ribbon (b),  while in the flat system (a)
the bands remain spin-degenerated. 
The only noticeable difference in the flat case is a small shift in the bands at the $\Gamma$ point, as it can be observed in the zoom of Fig.  \ref{3_71}. Otherwise, the effect of SOC is negligible. However, a large splitting is found in the curved ribbons, greater for the conduction bands. 
This is due to the interaction of edge states with higher-energy bands (see zoom in Fig. \ref{3_71} (b)), which in fact are completely hybridized due to curvature. Throughout most of the BZ, one of the spin-split bands has an upward shift in energy, whereas the other band undergoes a downshift. Thus, the bands without SOC mostly lie between the spin-orbit-split bands, as it can be seen in Fig. \ref{3_71} (b), especially in the zoom. 

Notice that 
GNRs can be made metallic because of the curvature, as seen in Fig. \ref{3_71}: the gap observed in the flat 1(7,1) ribbon (panel (a))  is 
still present at $\Gamma$ in the curved ($\varphi=2\pi$) geometry, but in this latter case the ribbon is metallic due to the band bending produced by curvature-induced hybridization (panel (b)).

Although the spin is no longer a good quantum number, the expectation value of the spin operator shows that in the flat geometry the total spin is normal to the ribbon. Curvature provokes the appearance of a small component in the in-plane direction which increases with curvature.

\begin{figure}[htbp]
\includegraphics*[width=\columnwidth]{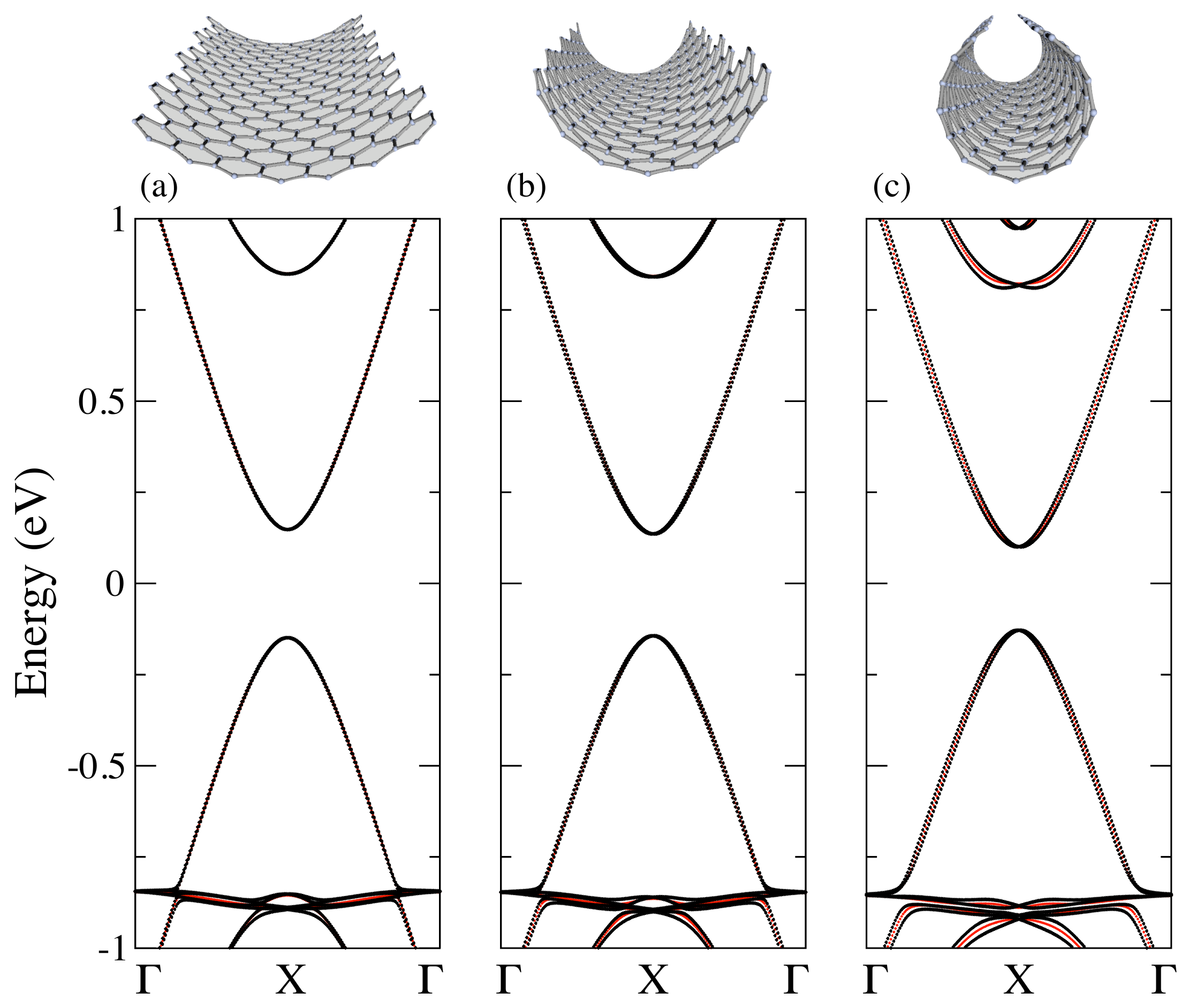}
\caption{(Color online) Electronic structure of the 1(3,2) GNR
 with different curvatures: (a) $\varphi=0.5\pi$
(b)  $\varphi=\pi$, and (c) $\varphi=1.8\pi$. The bands with SOC are shown in black; bands without SOC are in red (gray).}
\label{32curv}
\end{figure}

 Fig. \ref{32curv}  illustrates the effect of curvature. It presents the band structure of the 1(3,2) ribbon for three bending angles, 
namely, (a) 0.5$\pi$; (b) 1.2$\pi$, and  (c) 1.8$\pi$, with (black dots) and without (red/gray dots) SOC.The curved geometries are shown above each band panel. 
In a wide $M(3,2)$ ribbon there are four edge bands at 0 eV  extending  from $\frac{2}{3}\Gamma X$ to $X$.\cite{JA10} In the case depicted in Fig. \ref{32curv} a large gap opens between the occupied and unoccupied edge bands due to size effects, which we discuss later on. 
There is a general increase of the band splitting with growing curvature, as expected, due to the increment of the $\sigma$-$\pi$ hybridization produced in the curved ribbons, analogous to the effect predicted\cite{CLM04,CLM09} and experimentally measured in CNTs.\cite{KIRM08} Moreover, band splitting in GNRs is anisotropic, band- and $k$-dependent, as also found in CNTs. 

\subsubsection{Edge modification}

As discussed above, planar GNRs with dissimilar edges also lack inversion symmetry. Different edges can be achieved either by adding or removing atoms in a symmetric edge ribbon, or by altering the bond lengths at one edge of a symmetric GNR. We have explored the magnitude of this effect, calculating the changes in the band structure of a planar 1(5,2) modified ribbon. We have considered two types of modifications:  two atoms of one edge have been removed, and the bond length of the zigzag atoms at one edge has been changed 10\%. With both modifications SOC breaks the spin degeneracy of the band structure; however, the splitting is two orders of magnitude smaller than that achieved by the effect of curvature. Thus, in what follows, we concentrate on the curvature mechanism as a means to break inversion symmetry.

\subsection{SOC effects and width of the ribbon}

\begin{figure}[htbp]
\includegraphics*[width=\columnwidth]{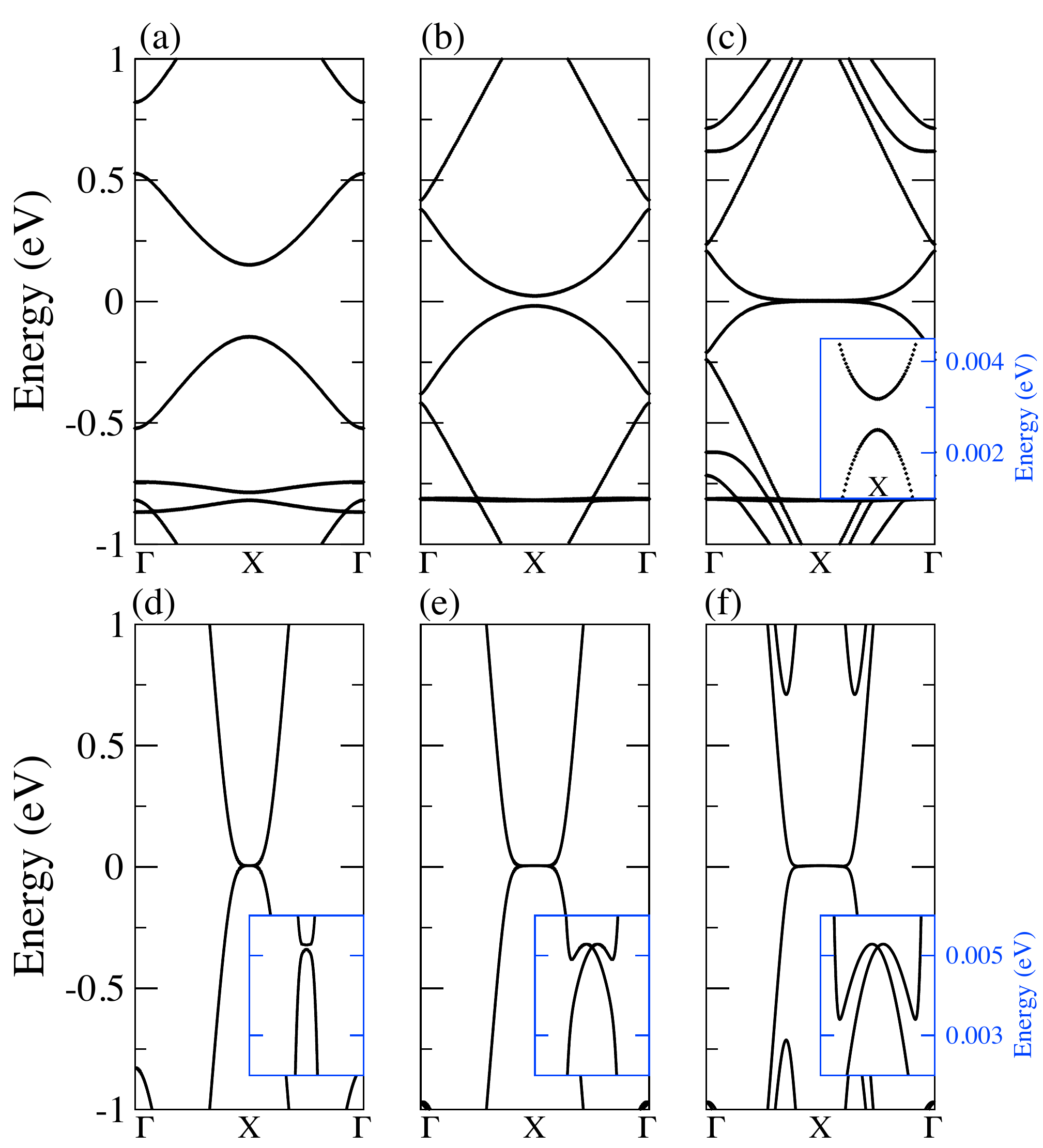}
\caption{(Color online) Band structures for flat ribbons calculated with
the $sp^3$ basis set, including the SOC term for the following GNRs:
(a) 1(5,2), (b) 2(5,2), (c) 4(5,2), 
(d) 2(1,0), (e) 4(1,0), and (f) 8(1,0). 
Insets show zooms of the bands near $E_F$.}
\label{52_zz}
\end{figure}

In order to explore the interplay of SOC and the width of the ribbon we consider first flat GNRs, since in curved geometries they could be masked by other effects,  such as hybridization. 
We have performed calculations for different chiralities, verifying that there is a gap in planar chiral nanoribbons that decreases with increasing width.
Panels (a) to (c) of Fig. \ref{52_zz} demonstrate 
this effect 
for the $M(5,2)$ GNRs. The 1(5,2) ribbon has a substantial gap, around  0.4 eV, while the flat bands around $E_F$ for the 4(5,2) ribbon in panel (c) are clearly identified as edge bands for their dispersionless character near the BZ boundary  X. Nonetheless, the gap can be discerned in the inset of Fig. \ref{52_zz} (c), as stated above. 

Comparison with high-symmetry zigzag ribbons of similar widths shows a striking difference. 
Figure \ref{52_zz} 
 demonstrates that zigzag ribbons have a negligible energy gap, around 0.1 meV for the narrowest case depicted [Fig.  \ref{52_zz} (d), $W=8.52$ \AA], 
 while for the 1(5,2) ribbon of similar width (8.87 \AA) the gap is around 0.4 eV. 

The gaps between edge bands in chiral ribbons are due to the stronger coupling
between edge states localized at each boundary. 
In zigzag 
ribbons the atoms at opposite edges 
belong to different sublattices, while in chiral ribbons boundary conditions at each edge mix the two sublattices,  coupling  
the states located at the two edges. 
This results in a band gap due to quantum size effects, without invoking electron-electron interactions. 

In flat ribbons, SOC induces a tiny shift of the energy bands; in zigzag ribbons, it turns the flat edge bands into dispersive ones. As inversion symmetry is preserved, all bands remain spin degenerated. 

\begin{figure}[htbp]
\includegraphics*[width=\columnwidth]{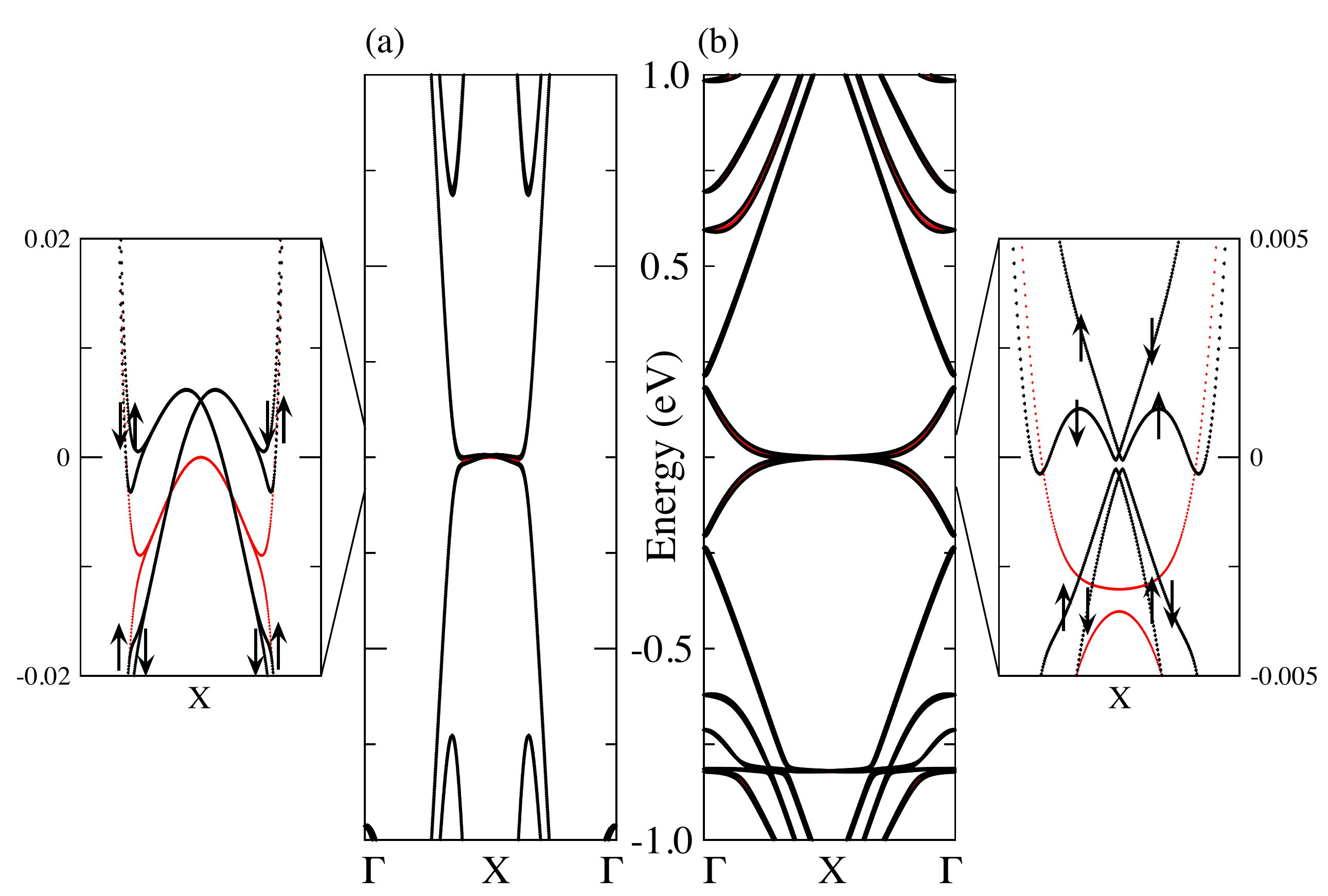}
\caption{(Color online) Band structures for curved ribbons calculated with
the $sp^3$ basis set, for the following GNRs: (a)  8(1,0), and  (b)  4(5,2). 
The bands with SOC are shown in black; bands without SOC are in red (gray).
Insets show zooms of the bands near $E_F$.}
\label{curv52zz}
\end{figure}

Now, we include curvature in order to enhance SOC effects and break spin degeneracy, as discussed in the previous Section. For narrow chiral ribbons the gap due to quantum size effects is rather large, as it can be seen in Figs. \ref{52_zz} (a) and (b),  so the effect of SOC is the aforementioned energy shift and, most importantly, the spin splitting of the bands.  We focus on the widest ribbons, namely, the 4(5,2) and the 8(1,0), with a smaller quantum size gap, and consider the same curvature radius for both ribbons, $R=6.274$ \AA, which yields an angle $\varphi=1.8 \pi$ and $1.73\pi$, respectively. 
For these cases,
the bands closer to $E_F$ are strongly deformed. 
 These happen to be edge states, so their behavior gives rise to a more interesting situation than in the large gap ribbons, as it is illustrated in Fig. \ref{curv52zz}. 

Fig. \ref{curv52zz} (a) shows the zigzag case, with a noticeable dispersion in the edge bands. The zoom shows that the bands with SOC are spin-split, with a crossing point slightly shifted with respect to that of the bands without SOC. Fig.  \ref{curv52zz} (b) shows the chiral 4(5,2) GNR; here besides the energy shift and spin-splitting of the SOC bands,  there is a slight displacement of the Fermi wavevector, which is no longer at X
Although small, SOC effects have important consequences for the transport properties of curved GNRs: spin-filtered channels arise due to the interplay of SOC and curvature, and for wider ribbons, even chiral GNRs present these spin-filtered channels in the low-energy region. 

\subsection{Chirality and spatial distribution of edge states}

\begin{figure}[htb]
\includegraphics*[width=\columnwidth]{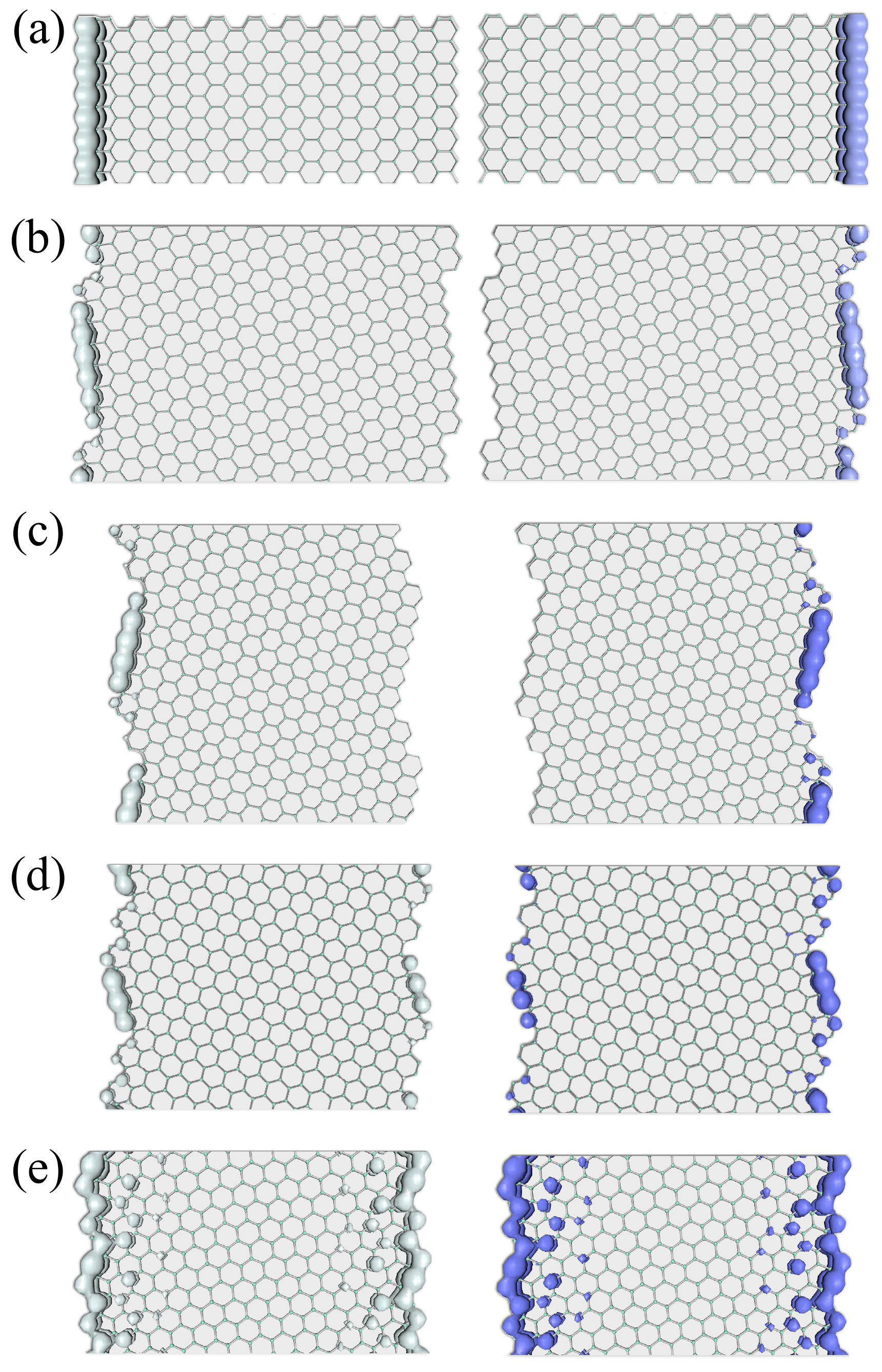}
\caption{(Color online) Probability densities corresponding to two states of the edge bands closer to $E_F$, near a high symmetry point, for several flat ribbons.  The spin polarization is plotted in white for spin up and blue (gray) for spin down. 
(a) Zigzag 10(1,0) ribbon, $k=0.95\,\Gamma$X; (b) chiral 4(7,1) ribbon, $k=0.10\,\Gamma$X; (c) chiral 1(7,2) ribbon, $k=0.95\,\Gamma$X; (d) chiral 4(5,2) ribbon, $k=0.10\,\Gamma$X;  (e) chiral 2(3,2) ribbon, $k=0.95\,\Gamma$X.}
\label{figdens}
\end{figure}
Edge states are among the most important features of GNRs.\cite{WTYS09,Waka_2012} 
It is interesting to explore how chirality affects the behavior of these states. 
 For the sake of simplicity, we  focus on flat geometries; generalization to curved geometries is straightforward.  In Fig. \ref{figdens} we present the electronic densities of edge states belonging to flat ribbons with different chiralities. The two states chosen correspond to the edge bands closer to the Fermi level, near the high-symmetry point to which the edge bands converge for large widths. Opposite spin polarizations are indicated with distinct colors; since we are dealing with flat ribbons, the spin direction is perpendicular to the plane of the ribbon. We choose a $k$ value slightly displaced from the symmetry point (either $\Gamma$ or the BZ boundary X) in order to avoid degeneracy due to time reversal invariance, and ribbons of similar width in order to compare size effects.  In pure zigzag ribbons, edge states are spin-filtered: each edge state has a well-defined spin orientation and it is located at one edge. This is illustrated in Fig. \ref{figdens} (a), which shows the square modulus of the wave functions for two edge states corresponding to a 10(1,0) zigzag nanoribbon, of width equal to 42.61 \AA. These two edge states of zigzag ribbons live in opposite sublattices, and their probability density is mostly confined to the atoms with coordination number 2 which constitute the geometrical edge. Fig. \ref{figdens} (b) shows the density for two edge states of a  4(7,1) ribbon of width 42.89 \AA, close in chirality to the zigzag case. 
 Similarly to the zigzag ribbon, the density of each edge is mostly located in one sublattice. It is not homogeneously distributed, being mostly in atoms with coordination number 2, although there is some appreciable weight in inner atoms close to the edge. For the 1(7,2) GNR (of width 34.88 \AA\ and $\theta=12.21^{\rm o}$) [Fig. \ref{figdens} (c)],  which has a larger chirality angle, the wave function extends more into inner atoms, especially close to the armchair part of the boundary. 

For chiralities closer to the armchair, the states have a nonzero density at both edges: panel (d) shows the two states close to $E_F$ for a 4(5,2) ribbon of width 35.48 \AA. These two states live at both edges simultaneously, with more inner atoms with nonzero density. This is more dramatic in panel (e) of Fig. \ref{figdens}, which depicts the edge states for a 2(3,2) ribbon of width 37.40 \AA . 
 As its chirality is closer to the armchair case, edge states have a greater penetration 
 into the inner part of the ribbon. In order to localize these states and obtain the quantum spin Hall phase, a larger value of the SOC constant ($\lambda\approx 4$ eV) is needed. This behavior has been found for armchair ribbons within the Dirac model\cite{ZS07} and the ETB model.\cite{LSM11} The SOC strength required increases with the chiral angle and decreases with the width of the ribbon. 

In curved ribbons, edge states keep their localized character, even for the maximum curvature. 
Besides the spin splitting (Fig. \ref{curv52zz}), 
there are two main differences:  the spatial localization
length is larger for the curved ribbon than for the flat case, and the spin direction changes from the direction normal to the ribbon surface, acquiring a component in the ribbon plane, as previously reported.\cite{LSM11, GPF11} 

\section{Discussion  and conclusions}

Our results show that the relation between the QSH edge states of graphene nanoribbons and both, the crystallographic orientation of the edges and the curvature of the GNR, allows us to control the spin and spatial localization of the ribbon edge states. Consequently, this may allow us to achieve an efficient electrical control of spin currents and spin densities in GNRs. 
Taking into account both spin and valley degeneracy, Bloch states in graphene are four-fold degenerate. SOC splits them into two Kramers doublets and, as Kane and Mele predicted,\cite{KM05} this turns graphene into a topologically non-trivial material. 
In zigzag edge ribbons, it has already been shown that at each edge spin-up and spin-down electrons move in opposite directions.\cite{LSM11,Chico_2012} Since backscattering in a given edge requires the reversal of spin it cannot be induced by spin-independent scatterers. Accordingly, edge states 
in zigzag GNRs 
are topologically protected and hence the conductance of the edge states is quantized.  However, in flat chiral-edge GNRs of finite width and for realistic values of the SOC strength, there is a non-zero probability of having electrons moving in opposite directions with the same spin polarization at a given edge (see fig. \ref{figdens}). Therefore, intra-edge backscattering may occur, which affects the quantization of the conductance. In chiral-edge GNRs, spin reversal can be induced even by non-magnetic disorder and thus edge states do not present a robust conductance quantization. The appearance of backscattering does depend on the chirality angle, increasing for angles approaching the armchair limit.  
On the other hand, curvature breaks the inversion symmetry of the ribbons and Bloch states are spin-split. Electrons with the same spin and opposite propagating directions  in a given edge have different energies.  As a result, backscattering is not allowed and in curved chiral-edge GNRs, edge states behave as robust quantum channels. 
Therefore, these chiral ribbons present a magneto-mechanical effect: upon curving the ribbon, the spin channels are split in energy, thus allowing for spatially separated spin currents. 

Despite the weakness of SOC in these carbon systems,---on the scale of a few meV---, their effects in curved graphene and nanotubes are not negligible, due to the coupling of $\pi$ and $\sigma$ bands in curved geometries. 
Thus, although small, the effects discussed in this work may be physically relevant,  and since in graphene the position of the Fermi level can be adjusted with external gate voltages, the control of spin currents in GNR-derived devices could be possible.

In summary, we have shown 
that, in the presence of spin-orbit interaction, curvature breaks spin degeneracy in graphene ribbons. Flat nanoribbons with symmetric edges, either chiral or achiral, have spin-degenerate bands. This is due to the existence of spatial inversion symmetry in flat ribbons, which is broken in the curved cases. 
Furthermore, spin-orbit splitting  
is enhanced 
in curved ribbons due to 
the hybridization of the bands, absent in
flat samples. Other mechanisms to break inversion symmetry, such as edge modification, are much less efficient to remove spin degeneracy. 

We have also explored finite-size effects in  GNRs. 
We find that narrow chiral ribbons present a sizable gap, despite their having a zigzag edge component, whereas in pure zigzag GNRs of similar width
 the gap is negligible.  
We relate this behavior to the boundary conditions in chiral edges, which mix the two sublattices at each edge. 
Finally, we have studied the chirality dependence on the spatial localization of edge states. 
In narrow chiral ribbons, edge states have 
a nonzero density at both edges simultaneously, due to edge coupling. 
Due to the sublattice mixing produced by the chiral boundary conditions,  
 edge states 
have a larger penetration than those of achiral ribbons.  For wider curved ribbons, they behave as spin-filtered states, being localized at one edge.

\bigskip

\section{Acknowledgments}
 H. S. gratefully acknowledges helpful discussions with J. E. Alvarellos.
This work has been partially supported by the Spanish Ministries of
Science and Innovation (MICINN) and Economy and Competitiveness (MINECO) DGES under
grants MAT2009-14578-C03-03, PIB2010BZ-00512, FIS2010-21282-C02-02, FIS2011-23713, MAT2012-38045-C04-04, and  FIS2012-33521.


%

\end{document}